\documentclass[fleqn,usenatbib]{mnras}

\usepackage{newtxtext,newtxmath}
\usepackage{anyfontsize}

\usepackage[T1]{fontenc}

\DeclareRobustCommand{\VAN}[3]{#2}
\let\VANthebibliography\thebibliography
\def\thebibliography{\DeclareRobustCommand{\VAN}[3]{##3}\VANthebibliography}

\usepackage{graphicx}	
\usepackage{amsmath}	
\usepackage{xcolor}

\title[BEBOP V: Spectral Analysis for BEBOP Targets]{BEBOP V.  Homogeneous Stellar Analysis of Potential Circumbinary Planet Hosts}

\author[Freckelton et al.]{Alix V. Freckelton$^{1}$, $^{\href{https://orcid.org/0009-0007-1053-0004}{\includegraphics[scale=0.5]{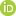}}}$\thanks{E-mail: AXF859@student.bham.ac.uk},
Daniel Sebastian$^{1}$ $^{\href{https://orcid.org/0000-0002-2214-9258}{\includegraphics[scale=0.5]{orcid.jpg}}}$, 
Annelies Mortier$^{1}$ $^{\href{https://orcid.org/0000-0001-7254-4363}{\includegraphics[scale=0.5]{orcid.jpg}}}$, 
Amaury H.M.J. Triaud$^{1}$ $^{\href{https://orcid.org/0000-0002-5510-8751}{\includegraphics[scale=0.5]{orcid.jpg}}}$,
\newauthor
Pierre F.L. Maxted $^{2}$ $^{\href{https://orcid.org/0000-0003-3794-1317}{\includegraphics[scale=0.5]{orcid.jpg}}}$,
Lorena Acu\~{n}a $^{3,4}$ $^{\href{https://orcid.org/0000-0002-9147-7925}{\includegraphics[scale=0.5]{orcid.jpg}}}$,
David J. Armstrong $^{5}$ $^{\href{https://orcid.org/0000-0002-5080-4117}{\includegraphics[scale=0.5]{orcid.jpg}}}$,
Matthew P. Battley$^{6}$ $^{\href{https://orcid.org/0000-0002-1357-9774}{\includegraphics[scale=0.5]{orcid.jpg}}}$,
\newauthor
Thomas A. Baycroft$^{1}$ $^{\href{https://orcid.org/0000-0002-3300-3449}{\includegraphics[scale=0.5]{orcid.jpg}}}$,
Isabelle Boisse$^{3}$,
Vincent Bourrier $^{6}$,
Andres Carmona $^{7}$,
Gavin A.L. Coleman $^{8}$ $^{\href{https://orcid.org/0000-0001-5111-8963}{\includegraphics[scale=0.5]{orcid.jpg}}}$,
\newauthor
Andrew Collier Cameron$^{9}$ $^{\href{https://orcid.org/0000-0002-8863-7828}{\includegraphics[scale=0.5]{orcid.jpg}}}$,
Pía Cortés-Zuleta $^{3,9}$ $^{\href{https://orcid.org/0000-0002-6174-4666}{\includegraphics[scale=0.5]{orcid.jpg}}}$,
Xavier Delfosse $^{7}$,
Georgina Dransfield $^{1}$
\newauthor
Alison Duck $^{10}$ $^{\href{https://orcid.org/0000-0002-4531-6899}{\includegraphics[scale=0.5]{orcid.jpg}}}$,
Thierry Forveille $^{7}$,
Jenni R. French $^{11}$ $^{\href{https://orcid.org/0000-0003-0825-4876}{\includegraphics[scale=0.5]{orcid.jpg}}}$,
Nathan Hara $^{6}$,
Neda Heidari $^{12}$ $^{\href{https://orcid.org/0000-0002-2370-0187}{\includegraphics[scale=0.5]{orcid.jpg}}}$,
\newauthor
Coel Hellier $^{2}$,
Vedad Kunovac $^{5, 13}$ $^{\href{https://orcid.org/0000-0001-9419-3736}{\includegraphics[scale=0.5]{orcid.jpg}}}$,
David V. Martin $^{14}$ $^{\href{https://orcid.org/0000-0002-7595-6360}{\includegraphics[scale=0.5]{orcid.jpg}}}$,
Eder Martioli $^{15,16}$ $^{\href{https://orcid.org/0000-0002-5084-168X}{\includegraphics[scale=0.5]{orcid.jpg}}}$,
James J. McCormac $^{5}$,
\newauthor
Richard P. Nelson $^{8}$ $^{\href{https://orcid.org/0000-0002-9687-8779}{\includegraphics[scale=0.5]{orcid.jpg}}}$,
Lalitha Sairam $^{1, 19}$ $^{\href{https://orcid.org/0000-0001-8102-3033}{\includegraphics[scale=0.5]{orcid.jpg}}}$,
Sérgio G. Sousa $^{17}$,
Matthew R. Standing $^{18,20}$ $^{\href{https://orcid.org/0000-0002-7608-8905}{\includegraphics[scale=0.5]{orcid.jpg}}}$,
Emma Willett $^{1}$  $^{\href{https://orcid.org/0000-0002-7831-1402}{\includegraphics[scale=0.5]{orcid.jpg}}}$
\\
$^{1}$ School of Physics \& Astronomy, University of Birmingham, Edgbaston, Birmingham, B15 2TT, UK \\
$^{2}$ Astrophysics Group, Keele University, ST5 5BG, UK \\
$^{3}$ Aix-Marseille Université, CNRS, CNES, Institut Origines, LAM, Marseille, France \\
$^{4}$ Max-Planck-Institut für Astronomie, Königstuhl 17,
D-69117 Heidelberg, Germany \\
$^{5}$ Department of Physics, University of Warwick, Coventry, CV4 7AL, UK \\
$^{6}$ Observatoire de Genève, Université de Genève, 51 Chemin Pegasi, 1290 Versoix, Switzerland \\
$^{7}$ Univ. Grenoble Alpes, CNRS, IPAG, 38000 Grenoble, France \\
$^{8}$ Astronomy Unit, School of Physical and Chemical Sciences, Queen Mary University of London, Mile End Road, London, E1 UK \\
$^{9}$ Centre for Exoplanet Science, SUPA School of Physics and Astronomy, University of St Andrews, North Haugh, St Andrews KY16 9SS, UK \\
$^{10}$ Department of Astronomy, The Ohio State University, Columbus, OH 43210, USA \\
$^{11}$ Department of Physics and Astronomy, University of Leicester, University Road, Leicester, LE1 7RH, UK \\
$^{12}$ Institut d'Astrophysique de Paris, UMR 7095 CNRS Université
Pierre et Marie Curie, 98 bis, boulevard Arago,  75014, Paris, France \\
$^{13}$ Centre for Exoplanets and Habitability, University of Warwick, Gibbet Hill Road, Coventry CV4 7AL, UK \\
$^{14}$ Department of Physics and Astronomy, Tufts University, 574 Boston Avenue, Medford, MA 02155,  USA \\
$^{15}$ Laborat\'{o}rio Nacional de Astrof\'{i}sica, Rua Estados
Unidos 154, 37504-364, Itajub\'{a} - MG, Brazil \\
$^{16}$ Institut d'Astrophysique de Paris, CNRS, UMR 7095, Sorbonne
Universit\'{e}, 98 bis bd Arago, 75014 Paris, France\\
$^{17}$ Instituto de Astrofísica e Ciências do Espaço, Universidade do Porto, CAUP, Rua das Estrelas, 4150-762 Porto, Portugal \\
$^{18}$ School of Physical Sciences, The Open University, Walton Hall, Milton Keynes. MK7 6AA. UK \\
$^{19}$ Institute of Astronomy, University of Cambridge, Madingley road, Cambridge CB3 0HA, UK \\
$^{20}$ European Space Agency (ESA), European Space Astronomy Centre (ESAC), Camino Bajo del Castillo s/n, 28692 Villanueva de la Ca\~{n}ada, Madrid, Spain
}

\date{Accepted 2024 June 01. Received 2024 May 24; in original form 2024 April 25}

\pubyear{2023}

\begin{document}
\label{firstpage}
\pagerange{\pageref{firstpage}--\pageref{lastpage}}
\maketitle

\begin{abstract}
Planets orbiting binary systems are relatively unexplored compared to those around single stars. Detections of circumbinary planets and planetary systems offer a first detailed view into our understanding of circumbinary planet formation and dynamical evolution. 
The BEBOP (Binaries Escorted by Orbiting Planets) radial velocity survey plays a special role in this adventure as it focuses on eclipsing single-lined binaries with an FGK dwarf primary and M dwarf secondary allowing for the highest-radial velocity precision using the HARPS and SOPHIE spectrographs. 
We obtained 4512 high-resolution spectra for the 179 targets in the BEBOP survey which we used to derive the stellar atmospheric parameters using both equivalent widths and spectral synthesis.
We furthermore derive stellar masses, radii, and ages for all targets. With this work, we present the first homogeneous catalogue of precise stellar parameters for these eclipsing single-lined binaries.
\end{abstract}

\begin{keywords}
techniques: spectroscopic -- binaries: eclipsing -- binaries: spectroscopic
\end{keywords}


\section{Introduction}

Circumbinary exoplanets orbit around both stars of a close binary system. The first confirmed systems have been discovered in the last decade using space-based transit observations e.g. Kepler-16\,b \citep{kepler-16b} and TOI-1338\,b \citep{kostov2020}.
Radial velocity (RV) detections of circumbinary planets have proven difficult because of stellar contamination, which has limited the RV precision that can be obtained in double-lined binaries \citep[e.g.][]{Konacki2010}. Nevertheless, \citet{triaud22} confirmed the radial-velocity signal of the circumbinary planet Kepler-16\,b. This was possible because the host binary is composed of a solar-type main-sequence star and a low-mass M-dwarf companion. This combination allowed us to treat the primary star spectroscopically as a single-lined star, for which achieving the necessary $\rm m\,s^{-1}$ precision to detect planetary signals is routinely done \citep[e.g.][]{Faria2022}.

The BEBOP survey for circumbinary planets (Binaries Escorted by Orbiting Planets) focuses on 
the EBLM project (Eclipsing Binaries with Low-Mass companions). Root-mean-square (RMS) scatter in residuals after removing the RV signal caused by the binary can reach down to about $3\,\rm m\,s^{-1}$ \citep{Standing2022}.
The survey has been designed as a blind all-sky survey and its sample was constructed from EBLM binaries detected through transit surveys \citep[e.g.][]{triaud2013,triaud2017,vonBoetticher2019,lendl2020}. Recently, BEBOP made its first discovery of a circumbinary planet solely based on RV measurements, EBLM J0608-59/TOI-1338/BEBOP-1\,c \citep[][BEBOP IV]{standing2023}.

Planet parameters, as measured through the transit or RV method, are inherently relative to their host star properties and such properties are often inferred from stellar evolution models \citep[e.g.][]{models}. Systematics introduced from different models are thus carried forward to analyses of the other planetary bodies in the system. It is therefore important to use a homogeneous and reliable set of stellar parameters for our BEBOP target stars to minimise any systematics in stellar as well as planetary parameters and ensure the statistical validity of the full survey. Several similar efforts have been done for other surveys \citep[e.g. ][]{sousa2011, Bucchave2014}. 

Throughout the BEBOP survey, we have already assembled a large archive containing thousands of high-resolution spectra of the primary FGK main-sequence stars. These spectra are first and foremost used to build up our RV time series, but they can also be used to measure stellar atmospheric parameters. 
Furthermore, thanks to the relative brightness of this sample ($\rm V \sim 8-13\,mag$) precise stellar parallaxes from Gaia DR3 \citep{GaiaCollaboration2016a,Gaia2023} as well as broadband photometry from the 2MASS \citep{2MASS} and AllWISE \citep{Cutri14} surveys are available for all targets. This allows us to use these in combination with our spectroscopic parameters to derive homogeneous masses, radii and ages for all targets.

In this work, we homogeneously measure stellar parameters for BEBOP's primary stars, using the same methodology overall with consistent input data. We analyse high-resolution spectra to derive the effective temperature ($\rm T_{\rm eff}$), surface gravity ($\log\,g_\star$), metallicity ($\rm [Fe/H]$), and projected rotational velocity ($v\sin i_\star$) of each star. We then use these parameters to derive stellar physical parameters such as mass ($M_\star$), radius ($R_\star$), and stellar age. These homogeneous parameters are of interest to produce accurate secondary star masses and radii as part of the EBLM survey \citep{triaud2017}, and accurate circumbinary masses in the context of the BEBOP survey \citep{bebop_1}. The paper is structured as followed: In Section \ref{sample} we introduce the BEBOP sample and in Section \ref{obs} the spectroscopic data. Our analysis method is described in Section \ref{paws} with results in Section \ref{sec:results}. Finally, we conclude in Section \ref{conclusions}.

\section{The BEBOP sample}
\label{sample}
A total of 179 systems are analysed in this paper with a magnitude range $\rm m_v = 8.31$ to $\rm m_v = 12.96$ (see Figure\ref{mag_snr}). The sample is split into a Northern sample (93 systems observed with the SOPHIE spectrograph, see Section \ref{SOPHIE}), and a Southern sample (110 systems observed with HARPS, see Section \ref{HARPS}). A total of nine systems are common between the Northern and Southern samples, selected on purpose in order to compare the sensitivity of both instruments to circumbinary exoplanets. This particular sub-sample is used to cross calibrate the spectroscopic parameters produced by both instruments. 

\subsection{BEBOP-South Sample}

The BEBOP-South sample was defined first. All systems identified as part of the EBLM sample \citep[see][]{triaud2013,triaud2017} were considered. Those EBLM secondaries were identified as transiting exoplanet false-positives during the WASP survey \citep[Wide Angle Search for Planets;][]{pollacco2006, triaud2011} thanks to the CORALIE spectrograph \citep{Udry2000}. All had received at least 13 spectra and some many more \citep[e.g.][]{bebop_1}. The BEBOP sample was selected to maximise radial-velocity precision, and minimise the contribution of the secondary stars. All visually identified double-lined binaries within the CORALIE spectra were removed. A Keplerian model was adjusted to all systems to obtain preliminary masses for the secondaries. All systems with binary period $P_{\rm bin} > 4.1~\rm days$ \citep[because no circumbinary has been found where $P_{\rm bin} < 5~\rm days$][]{martin18}, where the variance in the span of the bisector slope of individual spectra \citep[as defined in ][]{queloz2001} is below $177~\rm m\,s^{-1}$ , and where the full width at half maximum of the absorption lines is below $28~\rm km\,s^{-1}$ were kept (both to maximise radial-velocity precision and improve sensitivity to circumbinary exoplanets). This resulted in a sample of 56 eclipsing binaries, all expected to have a $\Delta V\rm mag >4$ between primary and secondary stars. 

In addition, we selected 22 systems identified as likely EBLMs by the KELT survey \citep[Kilodegree Extremely Little Telescope;][]{pepper2012,collins2018} with $\delta < +10^\circ$, $P_{\rm bin} > 5~\rm days$, with spectral types later than F4, and $V < 10.5$ for F-types, $V < 11$ for G-types and all K-types. Typically orbital parameters for these systems were more poorly determined than for the EBLM sample.

Finally, since {\it TESS} was launched \citep[Transiting Exoplanets Survey Satellite;][]{Ricker2014}, 32 new eclipsing systems consistent with EBLMs were added in 2021. These are typically brighter than the original EBLM sample, but usually without any prior radial-velocity information except in rare cases such as TOI-222 \citep[][]{lendl2020}. Their other properties, such as $P_{\rm bin}$, are consistent with the rest of the sample.

\subsection{BEBOP-North Sample}
The EBLM project's northern counterpart used SuperWASP to find candidates, and then SOPHIE to identify EBLM false positives \citep[e.g.][]{Gomez2013}. However, observations and classifications were not as systematic as in the South.  As such, the BEBOP-North sample was selected first from the KELT catalog \citep{collins2018}, cross-matched with SuperWASP/SOPHIE observations for confirmation. In addition all SuperWASP/SOPHIE false positives were reviewed selecting likely EBLMs from the eclipse depth and the absence of visible secondary eclipses. As in the South, only binaries with $P_{\rm bin} > 5~\rm days$, with spectral types later than F4 were kept. Only objects in the Northern hemisphere ($\delta > 0^\circ$) were selected, resulting in a sample of 120 systems. A first reconnaissance campaign was conducted on SOPHIE in 2018 to remove double-lined binaries and confirm the binary nature of each system, reducing the sample to 93 binaries. First all systems with $V >11.5$ were observed with the High Efficiency mode and all others in the High Resolution mode. After a few observations were collected, all systems with line widths $>15~\rm km\,s^{-1}$ were moved to High Efficiency mode since for those, spectral resolution is not as much of an issue.
\\

In both the Northern and Southern samples, systems were divided into a primary and a secondary sample. Typically, systems in the primary sample have more precisely measured RVs with the goal to collect of order 40 to 50 spectra and detect circumbinary planets. Systems in the secondary sample typically receive of order 10-15 measurements only.
\section{Observations}
\label{obs}

The data set analysed in this paper consists of 4512 high-resolution spectra obtained with the SOPHIE and HARPS spectrographs from 2013 to 2023, the majority of which (70\%) were observed with a 1800s exposure time. 

\subsection{SOPHIE spectroscopy}
\label{SOPHIE}
\begin{figure}
    \centering
    \includegraphics[width=0.5\textwidth]{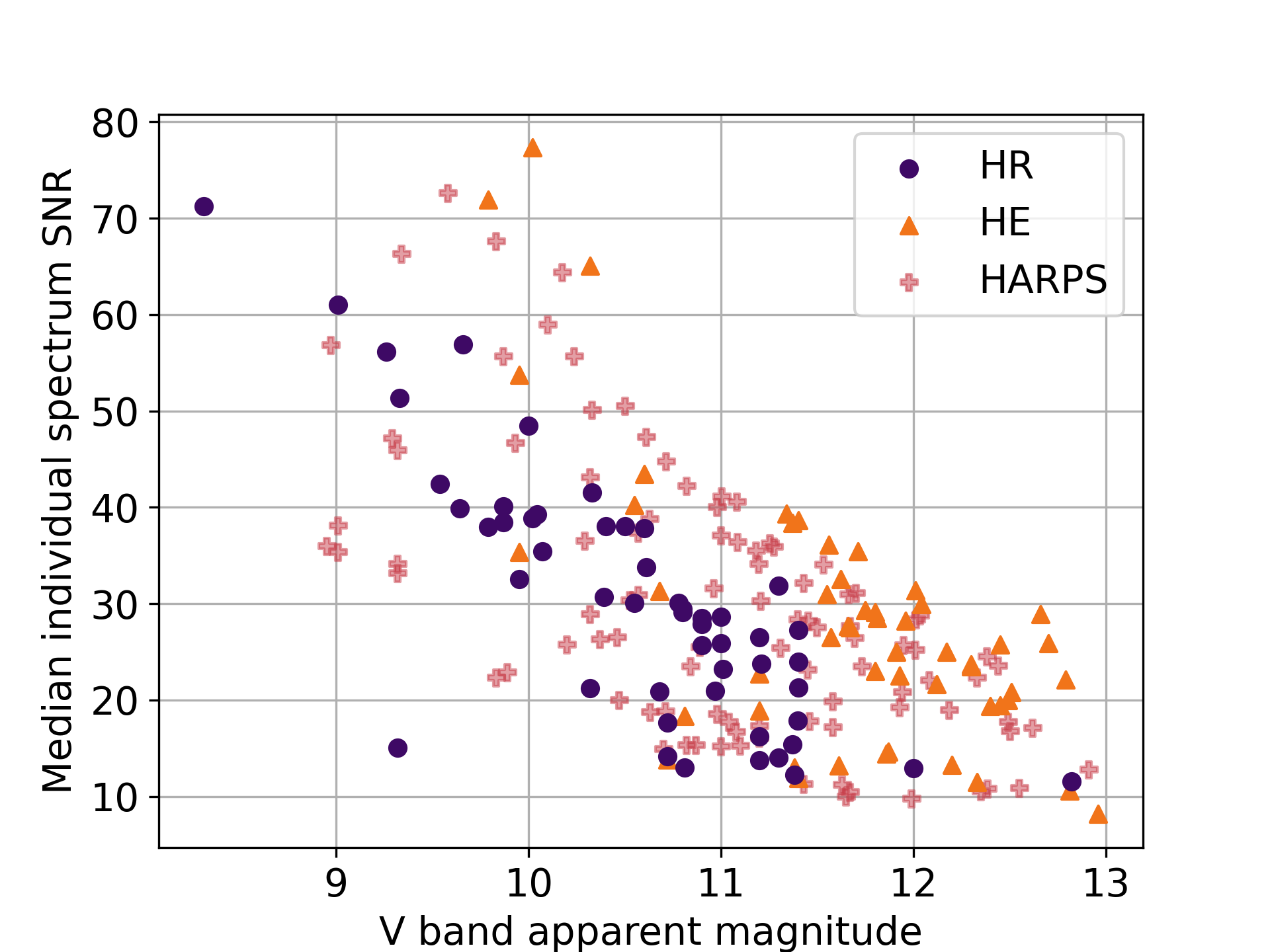}
    \caption{The median SNR of individual spectra is plotted against the apparent V band magnitude for each BEBOP sample target. The North Sample data is split into the two observation modes available with SOPHIE. The HR mode is shown as purple circles, and the HE mode as orange triangles. The South Sample, coming from HARPS, is shown as pink crosses. Note here that HARPS spectra have, on average, a shorter exposure time - this plot should not be taken as a representation of the instrumental performance.}
    \label{mag_snr}
\end{figure}

The SOPHIE échelle spectrograph \citep{SOPHIE,bouchy2009} is mounted on the $193~\rm cm$ reflector telescope at the Haute-Provence Observatory. SOPHIE has a wavelength range from $387.2~\rm nm$ to $694.3~\rm nm$. Two observation modes are available: high-resolution (HR) and high-efficiency (HE), respectively having resolutions of $R=75\,000$ and $R=40\,000$. 
Using the HE mode allows for a throughput increase equivalent to one magnitude. The versatility of the two modes is well demonstrated in the data set. Out of the 93 targets in the BEBOP Northern sample (taken using the SOPHIE spectrograph), 54 were observed in HE mode and 56 in HR mode (meaning 17 were observed in both modes). The median signal-to-noise ratio (SNR) achieved for individual spectra in the HE mode is 27, and the median SNR for the HR mode is 33. The sample is presented in Figure~\ref{mag_snr}. When considering the combined spectra used for analysis, the median SNR for the HR mode is 88, and 87 for the HE mode. 

All spectra are acquired as an \'echelle onto a CCD camera. The instrument is calibrated with Tungsten lamps at the start of every night to locate where each spectral order is, as well as to perform a flat field. Biases and darks are also obtained daily. In addition, Thorium-Argon lamps and a Fabry-P\'erot etalon are used to establish an accurate wavelength solution. A number of reference stars are observed every night in both the HR and HE mode to track the stability of the instrument \citep[typically of order $2~\rm m\,s^{-1}$;][]{bouchy2013,courcol2015,Hara2020}. Additional Fabry-P\'erot calibrations are obtained roughly every two hours throughout the night. 

SOPHIE operates with two fibres. All observations obtained for BEBOP use the {\tt objAB} mode where one fibre is on target and the other is on the sky so as to remove any contamination from e.g. Moonlight. Only one system was observed with the {\tt wavesimult} mode where the second fibre is instead illuminated by a Fabry-P\'erot etalon to produce a simultaneous calibration. This mode is usually reserved for systems where the most extreme radial-velocity precision is needed, which is not the case for the BEBOP stars. In the case of EBLM J0626+29 however, the sky fibre, which cannot be moved, coincided with another star. 

Using the calibration frames, each spectral order is extracted from the CCD by the SOPHIE Data Reduction Software (DRS), producing an {\tt e2ds} file. Each order is corrected for the instrumental blaze function and stitched together to create a one-dimensional spectrum, the {\tt s1d} files, which is what we use for our analysis. Each individual {\tt s1d}'s wavelengths solution is corrected to the barycentre of the Solar system \citep{bouchy2009,courcol2015}.

\subsection{HARPS spectroscopy}
\label{HARPS}
The HARPS spectrograph is  mounted on the ESO $3.6~\rm m$ telescope at La Silla observatory in Chile \citep{HARPS}. HARPS has a resolving power $R=115\,000$ over a wavelength range $378 \rm \: nm-691 \: nm$. Individual HARPS spectra used within this paper have a median SNR of 30. Observations and recording of spectra are done in a similar fashion to SOPHIE. Like SOPHIE, HARPS utilises two fibres, with one on the target and the other used as a calibration - either by use of a Th-Ar reference spectrum, or the sky background.

The data reduction for HARPS follows very closely the method outlined for SOPHIE. The main difference is that we only used the HR mode for observation with HARPS (called HAM). Like for SOPHIE, HARPS can either be used in {\tt objAB} or in {\tt wavesimult} mode. All BEBOP observations used the {\tt objAB} mode except for four systems identified by the {\it TESS} mission, because their brightness and spectral properties allowed us to reach a photon noise below HARPS's long term stability  \citep[around $1~\rm m\,s^{-1}$;][]{fischer2016} and a simultaneous calibration was necessary to make the best use of the instrument. HARPS is stable enough that neither reference stars nor intra-night calibrations are needed.

The radial velocity of the target spectrum can then be extracted by both the SOPHIE and HARPS reduction pipelines as the mean velocity from a Gaussian fitting to the cross-correlation function (CCF) profile. Out of all BEBOP targets, 6 were fitted with a K type mask by the HARPS and SOPHIE data reduction pipelines, with all others being fitted with G type masks.

Figure ~\ref{mag_snr} shows the HARPS spectra follow the same general trend as the SOPHIE data in terms of the SNR achieved for individual spectra. After coadding, the median spectral SNR is  126, surpassing that achieved by SOPHIE as outlined in Section \ref{SOPHIE}. All targets observed with HARPS were also observed using CORALIE, however the former was used due to the superior SNR achieved.

\section{Spectroscopic analysis}
\label{paws}
Two commonly used methods to retrieve atmospheric parameters via spectral analysis include the curve-of-growth equivalent widths (EW) \citep[e.g.][]{sousa2011,sousa2014} and spectral synthesis \citep[e.g.][]{valenti1996,adibekyan2012, tsantaki2018} methods. Alternative methods are also used extensively, such as the neural-network based Payne algorithm as used by the SAPP pipeline for the analysis of the PLATO core sample \citep{gent2022}.  The EW method uses the neutral and ionised absorption lines of only one element, resulting in a quick return of parameters. The atomic species used depends on the conditions of the star. For example, for young and/or active stars, it can be beneficial to use titanium (Ti) lines, as described by \citet{baratella2020}. However, for this work we employ the more commonly-used Fe I and Fe II lines, which are abundant in the spectra of main-sequence FGK stars.

On the other hand, spectral synthesis is more computationally intensive, iterating through parameters that synthesise a spectrum until the synthesised one matches the observed spectrum. Analyses of large data-sets would benefit from the speed of the EW method, however this method is unable to constrain the projected rotational velocity or macroturbulent velocity since a line's EW is conserved under these broadening parameters \citep[e.g.][]{sousa2011, santos2013}. To provide the most complete stellar information, the synthesis method must also, therefore, be implemented.

For the analysis of the BEBOP survey stars, we combine the strengths of both methods to provide a full and homogeneous analysis of the spectra of these FGK main-sequence stars. We utilise the speed of the EW method to provide excellent initial parameters for the synthesis method, which then runs much quicker. We use the {\tt iSpec} framework \citep{ispec1, ispec2} for our analysis. In this section we describe the details of our analysis and the tests done to ensure reliability of our parameters. In Appendix \ref{appendix:A}, we describe the public python pipeline, \texttt{PAWS}\footnote{\url{https://github.com/alixviolet/PAWS}}, we wrote around the {\tt iSpec} framework to perform our analysis.
\subsection{Data Preparation}
\label{datprep}

Barycentric velocity correction, made necessary due the the Earth's motion, is performed by the SOPHIE and HARPS reduction pipelines \citep{bouchy2009, lovis2007}, resulting in only needing to handle RV corrections in order to shift the BEBOP spectra into the lab frame. To perform the radial velocity correction, an atomic line mask developed from the NARVAL solar spectrum \citep{auriere2003} is used as a comparison template, representing the lab frame. A cross-correlation function (CCF) is then used to determine the radial velocity of the target star, which is then corrected for in the individual spectra to shift them into the lab frame. 

Consistent continuum normalisation is a crucial step to ensuring homogeneity throughout the analysis. Continuum flux in every spectrum is allocated as 1.0, with all spectral features, and therefore analysis, relative to this. To perform the normalisation, we use the \textit{fit\_continuum} function implemented in {\tt iSpec}. Continuum fluxes are found using a median filter of window size 0.05 nm, and maximum filter of 1.0 nm. Noise is identified by the median filter, then the maximum filter is used to block the fluxes in absorption lines. A model of the continuum is created by fitting a B-Spline of 2 degrees to the spectra, every $5~\rm nm$. The spectrum is then normalized by dividing all fluxes by this model.

Individual spectra from the same target are co-added prior to analysis to form one higher-SNR spectrum. Average flux and flux error at wavelength steps of $0.001~\rm nm$ are taken with a default range of $420-680~\rm nm$. Testing yielded no significant differences in results for varying wavelength steps. All BEBOP spectra were treated with the same wavelength range to ensure homogeneity. Spectra from the BEBOP sample are supplied without flux errors; we estimated these by dividing the flux at each pixel by spectral SNR provided in the FITS headers.

\subsection{Line List and Model Atmosphere}
\label{linelist}
A line list must be input during spectral analysis - this provides a subset of lines in the spectrum that will be used to determine the atmospheric parameters. To ensure homogeneity throughout the analysis, the same line list was employed for both the EWs and synthesis parts of our method. The line list created for the \texttt{SPECTRUM} code \citep{spectrum}, built on the NIST Atomic Spectra Database \citep{nist}, was chosen due to its proven success and versatility in FGK dwarf analysis \citep{ispec1}.

Absorption lines selected for analysis in the spectra are identified using a line mask. The line masks provided with iSpec contain atomic information inherently dependent on the spectral resolution of $R=47\,000$ for which they have been optimised. For use for the SOPHIE HE mode spectra, minimal modification was required for this line list, with only 27 lines removed due to consistently performing poorly in chi-squared testing - the wavelengths of these lines can be found in Table \ref{tab:lines}. HE mode spectra are more at risk of suffering from `blended' lines, in which the shallower and broader lines of the lower resolution spectrum may blend together.

To preserve the higher resolution achieved in the SOPHIE HR and HARPS spectra, we developed new line masks for this work using iSpec, which are publicly available on GitHub. For use with the HARPS spectra, we identified line masks using the HARPS-N solar spectra \citep{dumusque2021}, and the NARVAL solar spectrum at a resolution of $R=65\,000$ \citep{auriere2003} for the SOPHIE HR spectra. Individual abundances were then calculated for each line in these masks - where these were not within 0.05 of the accepted solar values, the masks were discarded to avoid lines that would not be suitable for atmospheric parameter determination. In the case of spectral synthesis, this constraint was extended to within 0.1 dex of solar abundances to allow for more lines to be available for synthesis.

Both analysis methods require a model atmosphere grid to be input. We chose to use the ATLAS9 set of model atmospheres \citep{atlas}. ATLAS is inclusive of a range of 4500 to 8750 K in T$\rm_{eff}$, 0.00 to 5.00 dex in  $\log\,g_\star$, and -5.00 to 1.00 dex in metallicity. Computation time is saved by the models only working with plane-parallel geometry atmospheres, which assume local thermodynamic equilibrium (LTE) and neglect 3D convection. This assumption breaks down for cold giant and super-giant atmospheres - although on the whole valid for FGK stars, it should be cautioned that biases may be introduced in Fe abundance and should be considered when abundances accurate to a few percent are desired, as described by \cite{bergemann2012}.

Studies such as those by \cite{cooke} often demonstrate extreme difficulty in constraining $\log\,g_\star$ from SOPHIE HE mode spectra, with that particular study reporting their $\log\,g_\star$ value with a significant error of 0.22 dex. Not only is such uncertainty seen in results from lower-resolution spectra, studies such as those by \citet{torres2012} also reveal $\log\,g_\star$ determination to be highly problematic in spectra of $R = 46 000, 48 000, 51 000,$ and $68 000$. Careful consideration was paid throughout this analysis to the fact that the spectroscopic $\log\,g_\star$ values derived for the BEBOP targets may be less reliable than those obtained via other methods such as photometry or via independently derived stellar mass and radius.

\subsection{Analysis Process}
\label{analysis}

Using either the EW or synthesis methods can produce reliable results. In this section, we discuss a robust approach that allows us to adopt both methods in an homogeneous way. Atmospheric parameters are first derived via the EW method, however due to this method's inability to constrain $v\sin i_\star$ and $v_{\rm mac}$ this does not obtain the full set of desired parameters. These results are thus used as initial parameters for the synthesis method, which derives the final set of atmospheric parameters.

\subsubsection{Equivalent Widths}
\label{EW}

We first use the EW method, employing only the Fe I and Fe II lines from the line list. We fit a Gaussian profile to each spectral line separately using \texttt{ARES} \citep{sousa2015}. The equivalent width is defined by taking a rectangle with a height equal to that of the continuum, and varying the width of it until the rectangular area is equal to that of the line under the continuum. The WIDTH radiative transfer code \citep{width} then derives individual abundances for these Fe lines based on a set of initial atmospheric parameters - in this work, these were set to the solar values collated in \citet{ispec2} since the BEBOP targets were chosen to be FGK dwarfs. 

Through a minimisation procedure, stellar parameters are then varied with the best fitting parameters ensuring ionisation and excitation balance. We fit for T$\rm_{eff}$, $\log\,g_\star$, and [Fe/H].  $v_{\rm mic}$ is estimated using the \textit{estimate\_vmic} function included in {\tt iSpec}; this relation depends on T$\rm_{eff}$, $\log\,g_\star$, and [Fe/H], and was derived using the results of GES (Gaia-ESO Survey) UVES data release 1 \citep{jofre2014, ispec1}.

\subsubsection{Spectral Synthesis}
Unlike the EW method, spectral synthesis uses every line in our line list. The WIDTH radiative transfer code is not able to perform spectral synthesis, so this part of the pipeline calls upon the SPECTRUM code \citep{spectrum}, chosen due to proven speed and reliability in the analysis of FGK dwarf stars by \citet{ispec2}.  Stellar atmospheric parameters are used with SPECTRUM to generate a synthesised spectrum. 
Parameters are iterated and used in conjunction with a minimisation algorithm to determine the optimal fit of the synthesized spectrum to the observed one. T$\rm_{eff}$, $\log\,g_\star$, [Fe/H] and $v\sin i_\star$ are fitted for, whereas $v_{\rm mic}$ and $v_{\rm mac}$ are calculated using \textit{estimate\_vmic} and \textit{estimate\_vmac}, with the relation for $v_{\rm mac}$ again based upon the GES UVES results as described for $v_{\rm mic}$ in Section \ref{EW}. 

The synthesis process is sensitive to its initial conditions. We therefore use the EW method to set reasonable estimates for the first iteration. This ensures that the synthesis begins in a parameter space that reflects the observed spectrum, saving considerable time compared to if beginning from a solar input for all targets. We set the maximum number of iterations to 6; \cite{ispec1} details how this is the optimal number as more iterations can cause metallicity dispersion to be favoured disproportionately compared to other parameters. The errors on derived parameters are calculated using the covariance matrix generated by the least-squares fitting \citep{ispec1}. In the case of T$\rm_{eff}$, the precision errors reported by iSpec are smaller than the expected accuracy of model atmospheres. To compensate for this, the errors are inflated by adding 100 K in quadrature to better reflect the uncertainty \citep{Tayar2022}.

\subsection{Handling Low SNR Data}
\label{lowsnr}
Since the SNR of spectra is known to be a critical aspect in stellar analysis, we analysed the individual results of using varying SNR spectra for the same target. Such testing was aimed primarily at producing an SNR filter value, under which spectra are not used. \cite{porto} state that 90$\%$ of their spectra surpass an SNR of 200 for their equivalent widths analysis; suggesting this filter would mainly be in place for the first step of our method. We made use of the target EBLM J0002+47, with the primary star being a slow-rotating F-type dwarf. The target has 33 spectra available with a large range of SNR, observed using the SOPHIE HR mode. Individual spectra are not combined as part of this testing; instead the EW and synthesis methods are applied separately using solar input parameters to each spectrum. 

\begin{figure*}
    \centering
    \includegraphics[width=\textwidth]{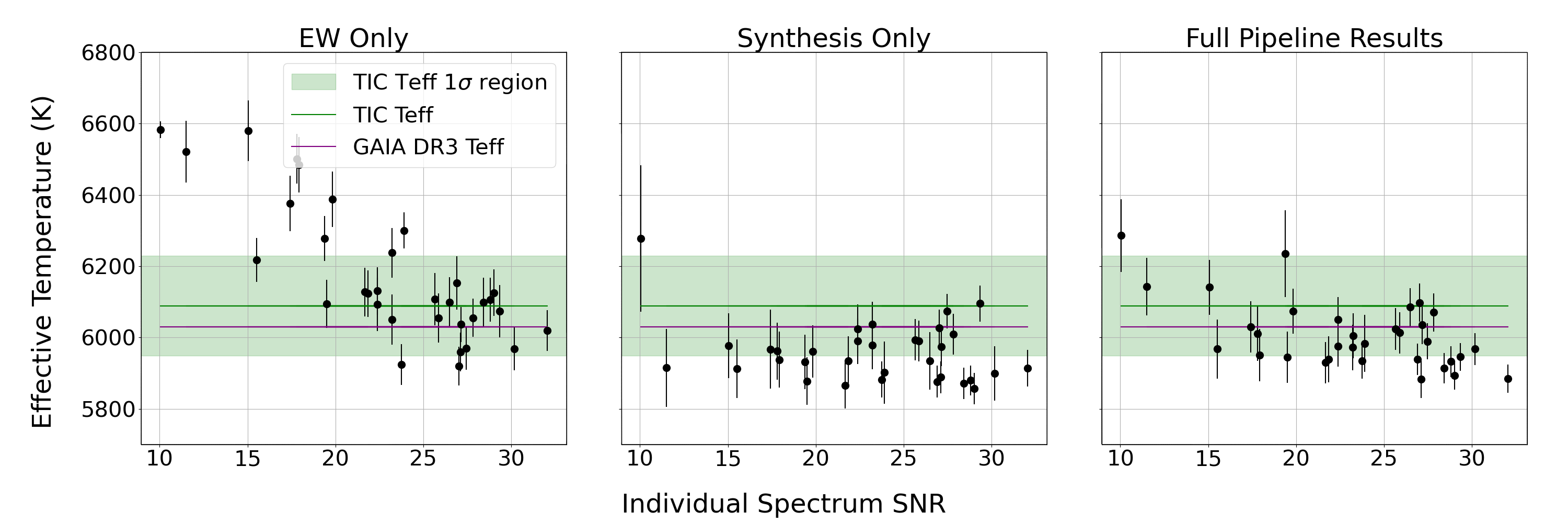}
    \caption{Across all three panels, the TIC T$\rm_{eff}$ value for J0002+47 is represented by the green horizontal line, with its error bars shown as the green shaded region. The purple horizontal line represents the Gaia DR3 T$\rm_{eff}$, supplied without errors. a)  T$\rm_{eff}$ derived from varying SNR spectra of J0002+47 using only the equivalent widths method.   b) Effective Temperature derived from varying SNR spectra of J0002+47 using only the synthesis method.  c)  T$\rm_{eff}$ derived from varying SNR spectra of J0002+47 using both methods subsequently.}
    \label{threepanel}
\end{figure*}

The results of using only the EW method are shown in Figure \ref{threepanel}a, shown as the derived T$\rm_{eff}$ for each individual spectrum plotted against its SNR. Included within the plot is a comparison to literature T$\rm_{eff}$ values for the target, with the TESS Input Catalogue (TIC) value \citep{tess} represented in green with the appropriate error range of 140 K, and the Gaia DR3 value represented in purple \citep{GAIA_1, GAIA_2, GAIA_3}. Although lower than the TIC value, the Gaia DR3 T$\rm_{eff}$ (shown as the purple horizontal line) falls within the TIC T$\rm_{eff}$ errors.  A very clear trend is displayed, showing a clear deviation from literature T$\rm_{eff}$ decreasing as spectral SNR increases, as one would expect. An SNR of 20 marks a significant cut-off, above which the majority of results lie within the accepted range of the TIC T$\rm_{eff}$ value. To ensure deviations such as those shown in this plot are kept to a minimum, individual spectra with SNR $<$ 20 were filtered out prior to analysis. The root-mean-square deviation (RMSD) from the TIC value using all results EW-only testing is 208 K, whereas after removing spectra with SNR $<$ 20 decreases the RMSD to 91 K - a significant improvement.  An additional concern is displayed in the lowest SNR spectrum of Figure \ref{threepanel}a having the smallest error in T$\rm_{eff}$; this suggests that the error derived for low SNR spectra by only the EW method does not accurately reflect the uncertainty of the value. Indeed, these uncertainties are statistical only and do not reflect all systematic effects involved in the parameter determination as they are not intended to be the final product of the method. Additionally, the uncertainties determined from only the EW method are highly sensitive to the flux errors supplied \citep{ispec1} - if these were estimated incorrectly, the uncertainties would be affected.

Figure \ref{threepanel}b shows the result of using only the synthesis method on the individual spectra of J0002+47. No obvious trend is observed in the T$\rm_{eff}$ values derived with respect to the spectral SNR. Instead, a systematic under-estimation of between 50 and 200 K below the TIC T$\rm_{eff}$ is displayed from the majority of spectra. This is likely to be influenced by the solar input parameters used by default with the synthesis method (i.e. when no input parameters are specified), with all results scattered within a much smaller region than in Figure \ref{threepanel}a. It is interesting that the synthesis method produced an effective temperature within error-bars of both literature values using the lowest SNR spectra available, however the lower SNR is reflected by a poorer synthesis fit and thus larger uncertainties. Again investigating the RMSD of the results, the synthesis-only testing has a RMSD of 155 K from the TIC value, greatly reduced compared to the 208 K from EW-only testing. Figure \ref{threepanel}b visibly shows much less scatter in the results than Figure \ref{threepanel}a; if we consider the RMSD from the mean of the results, rather than from the TIC value, a value of 83 K is achieved. This demonstrates the ability of the synthesis method to derive consistent parameters regardless of spectral SNR.

T$\rm_{eff}$ derived by spectral synthesis appears independent of spectral SNR, however indicates an underestimation bias.

Figure \ref{threepanel}c demonstrates the benefits of combining both methods. Where the synthesis method displays a bias to its input parameters, and the EW method has particularly poor performance at low SNR, combining the two gives a far lower dispersion of results. Returning to the metric of RMSD from the TIC T$\rm_{eff}$, using the entire pipeline returns an RMSD of 123 K, the lowest of the three tests. Additionally, the RMSD from the mean T$\rm_{eff}$ determined from the pipeline is 92 K, again showing strong consistency. In fact, all but one of the results agrees with the TIC values. Agreement is stronger with the Gaia DR3 effective temperature in Figure \ref{threepanel}c, as is also reflected in the higher SNR region of Figure \ref{threepanel}a. No metric is available to determine which value represents the most reliable result, so within this paper they are treated as equally likely. Given that the results in Figure \ref{threepanel}c agree largely with both values, there is no cause for concern or necessity to prove one value as more physically correct than the other.

In cases where the SNR of the combined spectrum is extremely low (< $\approx$ 50), we chose to skip the EW step and purely use the synthesis method for analysis. As shown in Figure \ref{threepanel}, the synthesis method is far less affected by low SNR spectra, whereas the EW method can produce results that differ hugely from expectations. 
We chose to only compare the effective temperatures as photometric metallicity may not be a reliable reference \citep{Morrell2019}.

\subsection{Handling Fast Rotators}

It is well-established that for fast-rotating stars ($v\sin i_\star\gtrsim 5 ~\rm km\,s^{-1}$) the reliability of the EW method is reduced due to the blending of spectral lines \citep{tsantaki2014}. To deal with this, we adopt the same approach as in Section \ref{lowsnr}, in which the EW method is skipped. The full list of targets for which the EW method was skipped is shown in Table \ref{tab:skippy}. The blending of lines by high rotational velocity manifests as an increased FWHM of the CCF described in Section \ref{datprep}, which is essentially the average shape of the spectral lines. The FWHM of the CCF can therefore be used to determine an estimate for the $v\sin i_\star$, as detailed by \citet{rainer2023}. Where the FWHM indicates that the star has $v\sin i_\star\gtrsim 5 ~\rm km\,s^{-1}$ (i.e. FWHM $ \gtrsim 20 ~\rm km\,s^{-1}$), only the synthesis method is used in analysis, with solar parameters as inputs, and the initial $v\sin i_\star$ set to that estimated from the FWHM.

\subsection{Testing on J2046+06}
\label{2046_test}
We used EBLM J2046+06 to test the output of our combined method, due to having recent and reliable literature parameters determined spectroscopically by \cite{eblmVIII}. Using 22 HARPS spectra combined to an SNR $\approx$ 300, their analysis uses ARES+MOOG, as described by \cite{sousa2014} and \cite{santos2013}. Differing from our method, they employ only the EW method using the ARES code \citep{sousa2007,sousa2015} together with the line list described in \cite{sousa2008}, Kurucz model atmospheres \citep{kurucz1993}, and the MOOG radiative transfer code \citep{sneden1973}. 

Both SOPHIE HR and HARPS spectra are available for EBLM J2046+06, allowing additionally for testing of continuity between different instruments.  

\begin{table*}
    \caption{Results of our analysis on EBLM J2046+06 from SOPHIE HR mode and HARPS spectra, compared to results from \citet{eblmVIII} and Gaia DR3 \citep{recioblanco2023}. }
    \label{J2046+06}

    \begin{tabular}{ccccc}
    \hline
         Source & $\rm T_{eff}$ (K) & $\log\,g_\star$ (dex) & [Fe/H] (dex) & $v_{\rm mic}$ (km/s) \\
         \hline
         \cite{eblmVIII} & 6302 $\pm$ 70 & 3.98 $\pm$ 0.11 & 0.00 $\pm$ 0.05 & 1.61 $\pm$ 0.05 \\
         HARPS & 6314  $\pm$ 114 &  4.03 $\pm$ 0.13   &  -0.10 $\pm$ 0.09 &  1.84 $\pm$ 0.03  \\
         SOPHIE &  6231 $\pm$ 105 &  3.92 $\pm$ 0.15 &  -0.11 $\pm$ 0.07 &  1.67 $\pm$ 0.06 \\
          Gaia DR3 GSP-Spec &  6149 $\pm$ 53 &  4.08 $\pm$ 0.04 &   -0.13 $\pm$ 0.04 & - \\
         \hline
    \end{tabular}
\end{table*}

\begin{figure*}
    \centering
    \includegraphics[width=\textwidth]{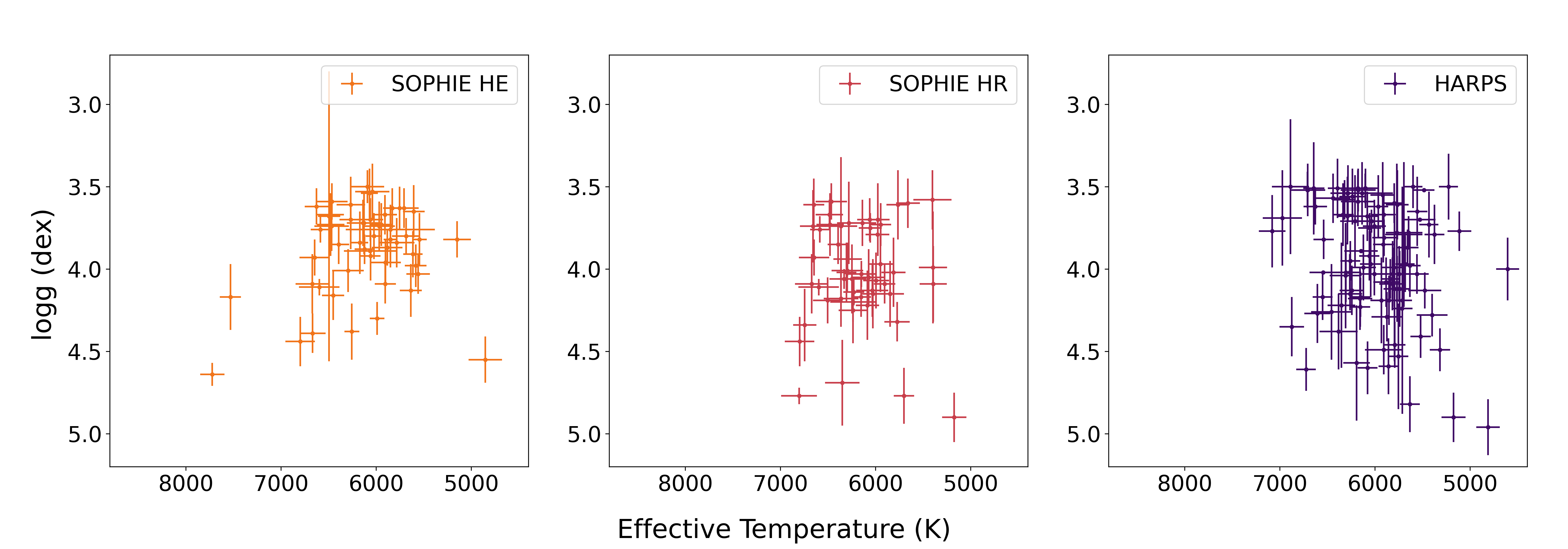}
    \caption{The BEBOP sample presented as a $\log\,g_\star$ vs T$\rm_{eff}$ diagram. The data are split into the respective source of the spectra, ordered as SOPHIE HE (left), SOPHIE HR (middle), and HARPS (right).}
    \label{loggteff}
\end{figure*}

Table \ref{J2046+06} displays the results of the pipeline testing on both SOPHIE and HARPS spectra, in addition to the parameters determined by \cite{eblmVIII} and Gaia DR3 GSP-Spec, determined using spectra from the Gaia's Radial Velocity Spectrometer (RVS). \citep{recioblanco2023}. The higher SNR and resolution achieved by HARPS is demonstrated as an advantage here, being closer to the values obtained by \cite{eblmVIII} than those from using the SOPHIE HR mode spectra. With 22 available spectra having an average SNR of 43, the SOPHIE combined spectrum has SNR 131, whereas the HARPS combined spectrum has an SNR of 273, from 27 spectra with average SNR of 57. Despite this, parameters within 2$\sigma$ are returned in both cases when compared to \cite{eblmVIII} in all cases except $v_{\rm mic}$. Table \ref{J2046+06} displays that atmospheric parameters derived from spectra from different instruments agree well with each other, including for lower SNR and resolution SOPHIE spectra. This strengthens the reliability of our methods. Concerning the results from Gaia DR3, our results, and those from \cite{eblmVIII}, good agreement is shown.

\section{Results}
\label{sec:results}

\subsection{Atmospheric parameters}
\label{results}

We analysed the spectra for all 179 targets in the BEBOP sample with the methods described in Section \ref{paws}. It took approximately 66 hours of computation time \footnote{Using 11th Gen Intel Core i7-1185G7@3.00GHz x 8.}. If a star was analysed using spectra from multiple spectrographs, the final adopted stellar parameters are calculated as an inverse-variance weighted average from the individual results. 

The results of the full sample are shown in Figure \ref{loggteff}, in the form of a $\log\,g_\star$ vs. T$\rm_{eff}$ diagram. SOPHIE HE mode, HR mode, and HARPS spectra are split into three separate panels on the same scale to demonstrate the dispersion of each. 

\subsubsection{Comparison with Gaia DR3}
\label{gaia}
\begin{figure*}
    \centering
    \includegraphics[width=\textwidth]{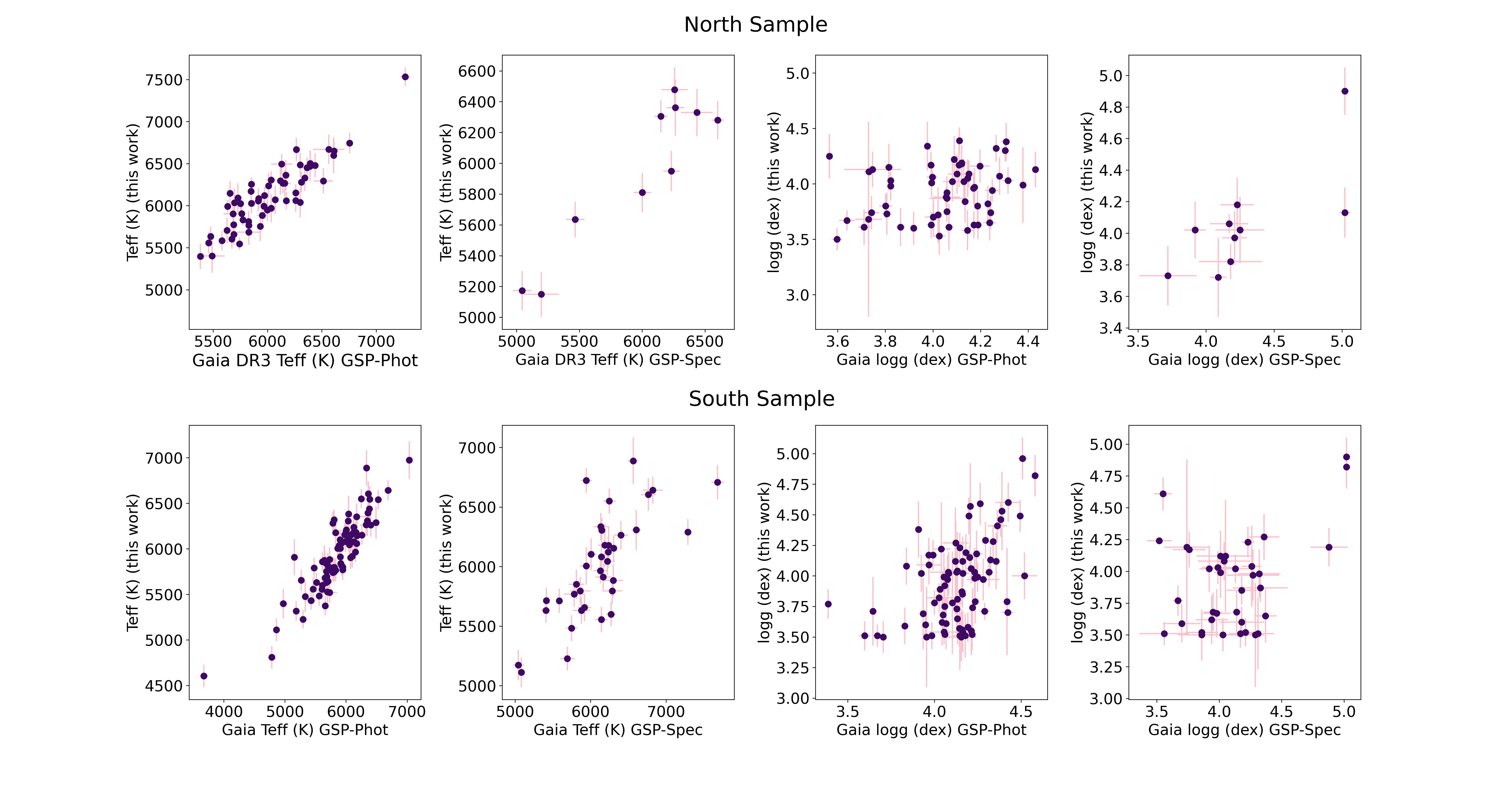}
    \caption{ Comparisons of our results (y axes) to Gaia DR3 parameters (x axes). Results from the North sample (both HE and HR mode) are the top two plots, and South sample the bottom two plots, with T$\rm_{eff}$ and $\log\,g_\star$ comparisons on the left and right respectively. }
    \label{lit_pipeline}
\end{figure*}

Gaia DR3 parameters were used in a comparison of our output to literature parameters due to providing results for the majority of our targets. Comparisons were performed for T$\rm_{eff}$ and $\log\,g_\star$, which were consistently available from both \texttt{GSP-Phot} \citep{andrae2023} and \texttt{GSP-Spec} \citep{recioblanco2023}. Figure \ref{lit_pipeline} shows our results versus the Gaia DR3 values, with \texttt{GSP-Spec} results limited to those with a \texttt{fluxNoise} quality flag of Flag 2 or lower, according to Appendix C of \citet{recioblanco2023}. These results were split into the North and South samples to check for potential instrumental biases. Although error bars are plotted for all points, it is important to consider that the uncertainties on the Gaia DR3 values represent precision values, whereas our T$\rm_{eff}$ were inflated in quadrature.

Figure \ref{lit_pipeline} shows good agreement between our results and the Gaia T$\rm_{eff}$ values. Considering the North sample, the RMSD between our results and \texttt{GSP-Phot} is 199 K; this is smaller than the mean error on our T$\rm_{eff}$ of 256 K. When comparing our results to those form \texttt{GSP-Spec}, the RMSD is 190 K, again being smaller than the mean error. For the South sample, the benefit of the higher resolution manifests in the RMSD between our results and \texttt{GSP-Phot} reducing to 184 K, with a mean uncertainty of 199 K. We see poorer agreement when comparing our results from the South sample to those from \texttt{GSP-Spec}, with the RMSD increasing to 266 K in this case.

Section \ref{linelist} discusses the difficulty in obtaining a reliable value of $\log\,g_\star$ from spectra, hence we did not expect to see concrete agreement when comparing our $\log\,g_\star$ to those from Gaia DR3. However, Figure \ref{lit_pipeline} does show that we do not determine any $\log\,g_\star$ values that would be unphysical for our targets, in addition to showing a general positive correlation between our results and results from Gaia DR3. Furthermore, we do not believe the discrepancies to be of great concern in terms of our final results, due to the extensive testing by \cite{mortier2014} that demonstrates the effect of a changing $\log\,g_\star$ to be insignificant on the derivation of other stellar parameters from spectroscopy.

\begin{figure*}
    \centering
    \includegraphics[width=\textwidth]{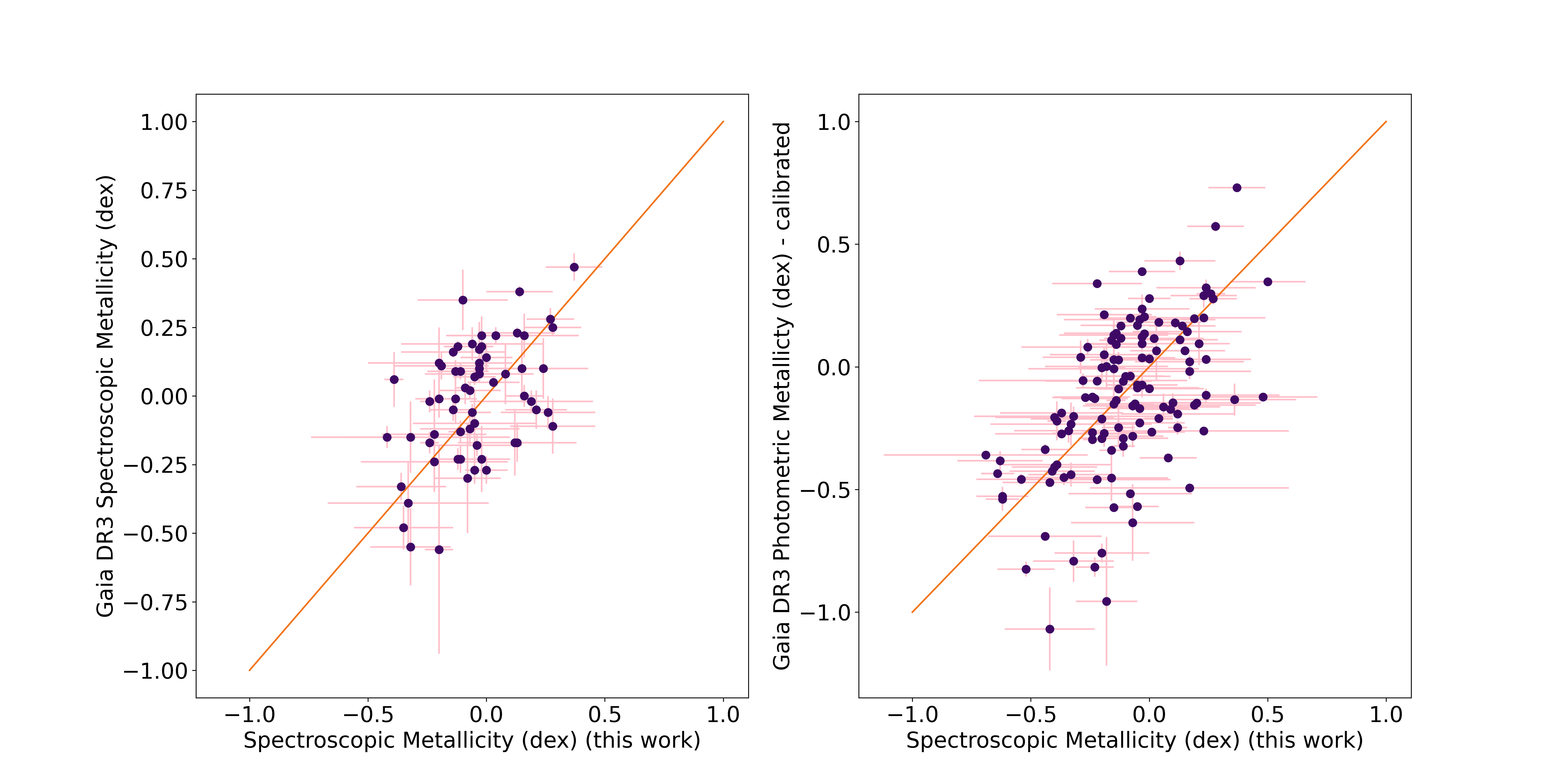}
    \caption{ Comparison of metallicties from this work (x axes) to Gaia DR3 \texttt{GSP-Spec} (left) and \texttt{GSP-Phot} (right).}
    \label{fig:gaia_feh}
\end{figure*}

As discussed by \citet{andrae2023}, Gaia \texttt{GSP-Phot} metallicity values suffer from a systematic underestimation.  To counteract this, we used the empirical calibration relation introduced by \citet{andrae2023} via the python package \texttt{gdr3apcal}\footnote{\url{https://github.com/mpi-astronomy/gdr3apcal}} to determine calibrated \texttt{GSP-Phot} metallicities. Figure \ref{fig:gaia_feh} shows a comparison of our results with both the results from \texttt{GSP-Phot} and \texttt{GSP-Spec}. Our results have a RMSD from the \texttt{GSP-Spec} values of 0.20 dex, increasing to 0.26 dex when we compare to the \texttt{GSP-Phot} values.

\subsubsection{BEBOP colour-magnitude diagram}

\begin{figure}
    \centering
    \includegraphics[width=0.5\textwidth]{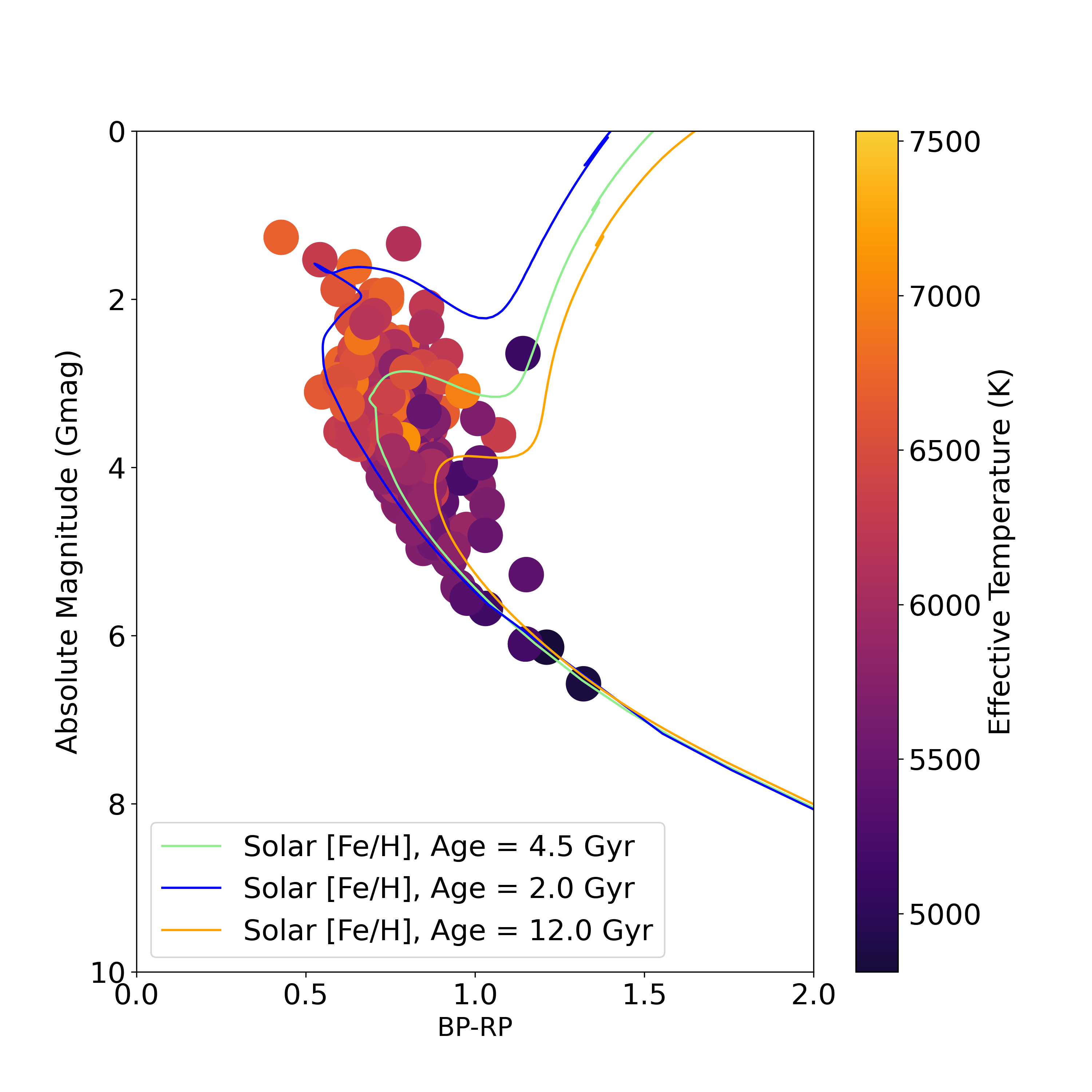}
    \caption{Colour-Magnitude diagram of the BEBOP sample, using Gaia DR3, with the colour representing our effective temperature. The cooler stars are represented by darker colours, becoming brighter as the effective temperature increases. The expected trend of the brightest, bluest stars (top left) being the hottest, and the dimmer, redder ones (bottom right) being the coolest is seen clearly here.}
    \label{cm_diagram}
\end{figure}

In Figure \ref{cm_diagram} we show t
he BEBOP sample in a colour-magnitude diagram using Gaia DR3 parallaxes, G band apparent magnitudes, and BP-RP colours \citep{GAIA_1, GAIA_2, GAIA_3}. No reddening was included so some scatter could arise from that, however, the stars have a mean distance of 336 pc, and reddening would thus be minimal overall. We add a colour-bar representing our derived effective temperature for each target. As one would expect the sample follows the main sequence with the hottest stars represented in this figure occupying the upper left of the colour-magnitude diagram, being the bluest and brightest from the sample. The reddest and dimmest stars are shown, as expected, to have the lowest effective temperatures.

\begin{figure}
    \centering
    \includegraphics[width=0.5\textwidth]{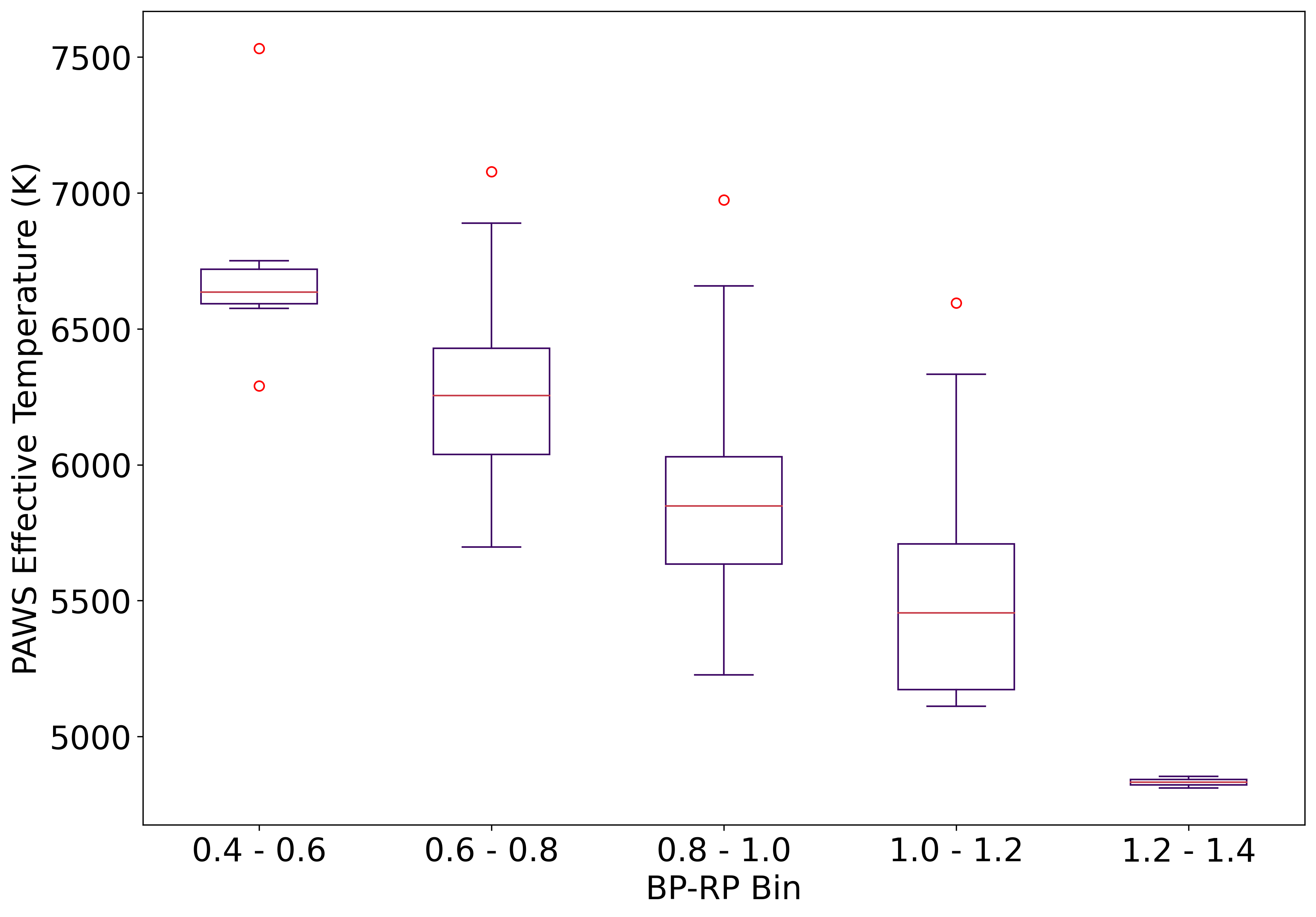}
    \caption{Box-and-whisker plot of the distribution of effective temperatures derived for targets in each Gaia DR3 colour bin, as taken from Figure \ref{cm_diagram}.}. 
    \label{box_outlier}
\end{figure}

Ensuring further that the effective temperatures decrease going from bluer to redder stars can present an excellent opportunity to reveal outliers. Figure \ref{box_outlier} separates the data displayed in Figure \ref{cm_diagram} into five bins of Gaia BP-RP colour, and uses a box-and-whisker plot to show the distribution of effective temperatures in each. Outliers from this are clearly displayed as the red points lying outside of the whiskers. Beginning with the 0.4-0.6 colour bin, the cool outlier is J0954-45.  Our results for this target show a $v\sin i_\star$ of 30.01 $\pm$ 16.01 $\rm km\,s^{-1}$ - such a high $v\sin i_\star$ could have resulted in blended lines that interfered with the analysis. There are no Gaia DR3 parameters available for J0954-45, hence we could not do a comparison here. The hotter outlier of the first bin corresponds to J1805+09, which has an exceptionally high $v\sin i_\star$ of 49.93 $\pm$ 9.30 $\rm km\,s^{-1}$, hence the target is also highly susceptible to blended lines. Our T$\rm_{eff}$ for this target was calculated to be 7532 $\pm$ 111 K; we can compare this to the Gaia DR3 T$\rm_{eff}$ of 7267 $\pm$ 49 K, keeping in consideration that Gaia DR3 errors are precision-only, whereas we have inflated our T$\rm_{eff}$ uncertainties in quadrature to reflect inaccuracies in the atmospheric models used for analysis. The Gaia DR3 result would also place the T$\rm_{eff}$ as an outlier in Figure \ref{box_outlier}. Further investigation of this target reveals the TIC radius reported to be 1.68 $\pm$ 0.07 R$_{\odot}$ - together with the pipeline-derived effective temperature for the HE mode of 7375 $\pm$ 180 K, \cite{mamajek} places this target in the range of late A to early F type stars. This indicates a possible unsuitability of J1805+09 as part of an FGK dataset. With an HE mode combined SNR of 65, additional observations of the target would be required to reach an SNR of 100 and reduce the uncertainty in parameters. 

Moving to the 0.6-0.8 bin, the single outlier displayed is J1258-58. This target has no Gaia DR3 results to compare to, and with the combined spectrum reaching an SNR of 136 we do not expect that the results are unphysical. The 0.8 - 1.0 colour bin also contains a single outlier, being J1916-04. Our result for the T$\rm_{eff}$ of this target is 6974 $\pm$ 206 K, which is in agreement with the Gaia DR3 T$\rm_{eff}$ of 7033 $\pm$ 32 K - this gives confidence in our results for this target. The final outlier in the 1.0 - 1.2 colour bin is J0525+26; given that the SNR of the combined spectrum is 42, we would require additional data to determine whether this target is an outlier due to its physical conditions or the quality of the data.

\subsection{Masses, Radii and Ages}
\label{m,r,a}

The atmospheric parameters of all BEBOP primaries allow us to derive stellar parameters such as mass and radius using stellar evolution models. We applied MIST isochrones \citep{Dotter16,Choi16} to interpolate the spectral parameters using the {\tt isochrones} package \citep{Morton15}. This interpolator utilises a multimodal nested sampler {\tt multinest} \citep{Feroz08,Feroz09,Feroz19} which allows to sample the input spectral parameters together with photometric and distance information. We used our derived $\rm T_{\rm eff}$ and $\rm [Fe/H]$, separating them for each instrument and instrumental mode, together with the Gaia DR3 parallaxes \citep{GaiaDR3_part1_22}, IR colours W1, W2, \& W3 from Allwise \citep{Cutri14}, and the cross matched NIR 2MASS \citep{2MASS} colours H, J, $\rm K_{s}$ from the same catalogue. We chose not to include a value for $\log\,g_\star$ as an input parameter due to being unable to determine its reliability. As we use a very precise parallax and a variety of magnitudes, it is also not a crucial parameter when fitting isochrones and evolutionary tracks. We fit for stellar mass, radius, age, distance and extinction ($\rm AV$). For each target, we sampled 1000 live points and extract the final parameters as median values from the posterior distributions with the errors representing the 16/84 percentile. Following the methodology outlined by \citet{Tayar2022}, noise floors of 5\% for masses, 4.2\% for radii and 20\% for ages were added to our uncertainties.

\begin{figure}
    \centering
    \includegraphics[width=0.5\textwidth]{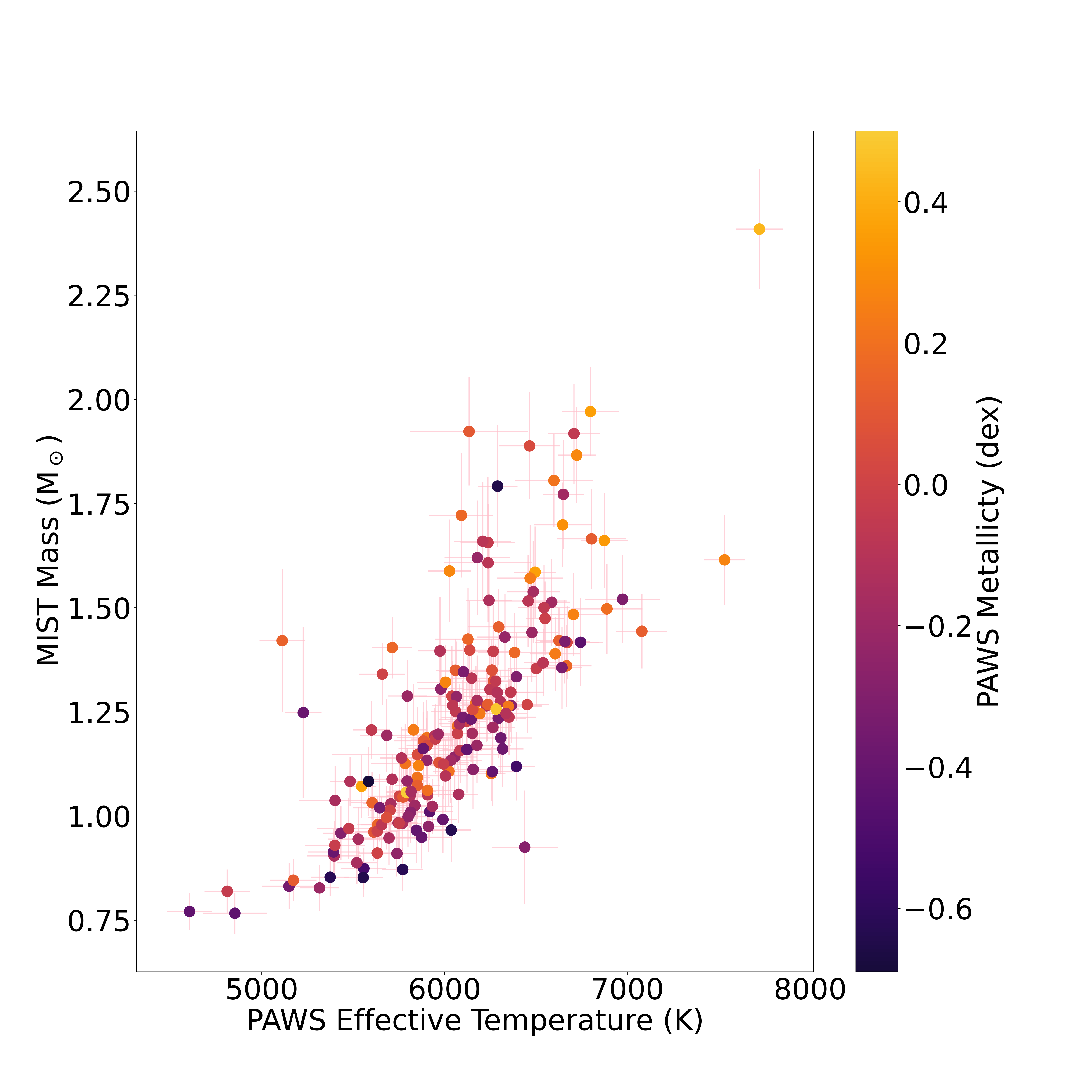}
    \caption{Masses obtained from MIST isochrones plotted against T$\rm_{eff}$, coloured by the metallicity of each target.}
    \label{mass_temp_feh}
\end{figure}

The resulting BEBOP primary mass-T$\rm_{eff}$ diagram, coloured by [Fe/H], is shown in Figure\,\ref{mass_temp_feh}. Each point in the plot is coloured by the metallicity of the target - this reveals an abundance of solar-and-higher metallicity targets. It is well established that the occurrence rate of giant planets is increased for metal-rich stars \citep{sousa2011}. Although not as numerous, many metal-poor targets can also be seen in this plot. These are interesting targets for the study of planet occurrence rates; \citet{mortier2012} demonstrate that giant planet frequencies around low metallicity stars may not be as diminished as initially expected. Future comparisons of the occurrence rates of giant planets found across the entire range of metallicity targets in the BEBOP sample with those previously studied around single stars would be highly beneficial to theories of planet formation. Having stellar mass values for primary stars in the BEBOP sample will allow for such comparisons to include good constraints on planetary masses.

\subsubsection{Spectroscopic and Isochronal Surface Gravities}
\begin{figure}
    \centering
    \includegraphics[width=0.5\textwidth]{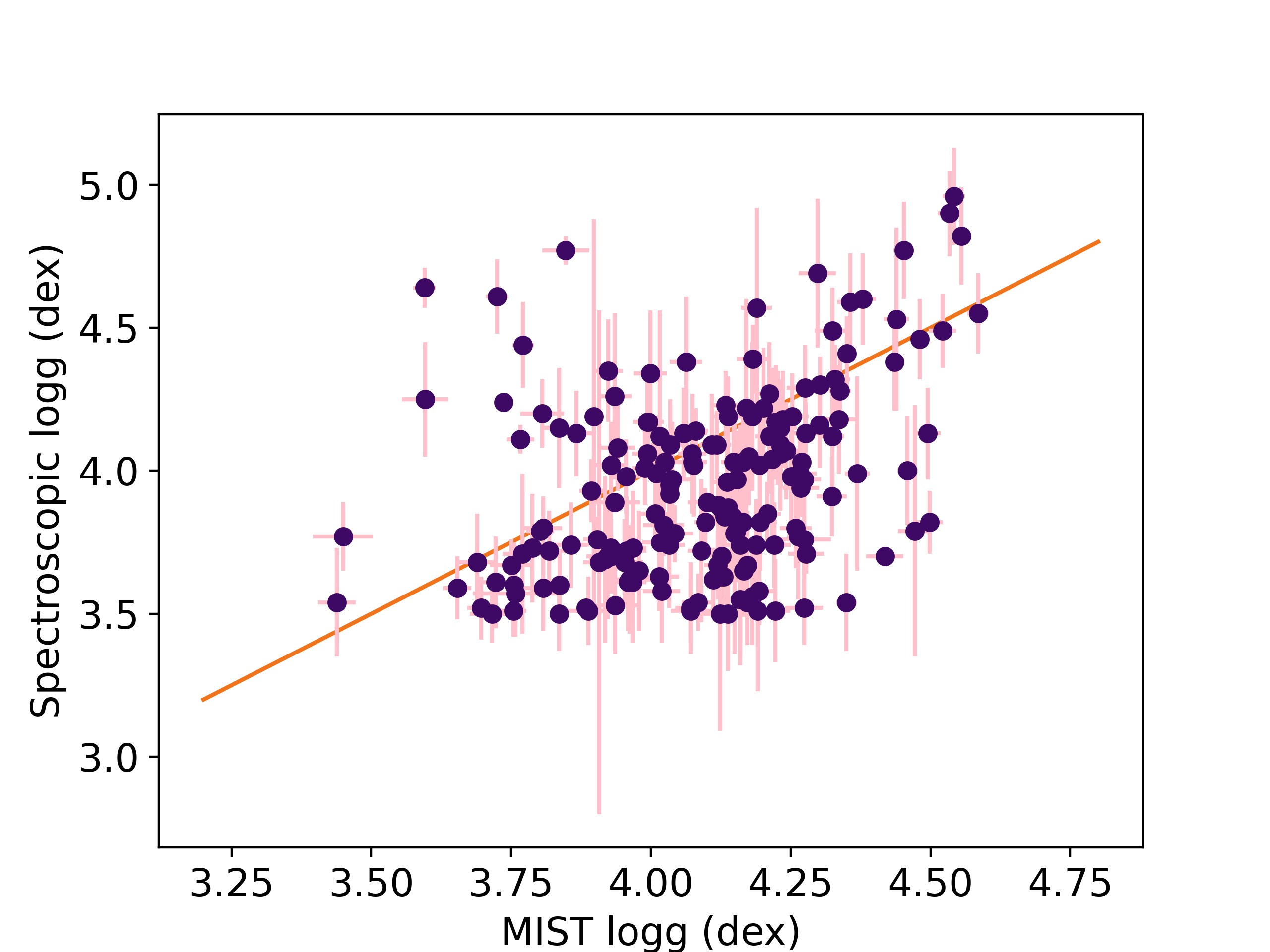}
    \caption{Spectroscopic $\log\,g_\star$ values compared with those inferred from MIST isochrones, both from this work.}
    \label{fig:logg_comp}
\end{figure}

 As highlighted in both Sections \ref{linelist} and \ref{gaia}, the spectroscopic $\log\,g_\star$ we derive in this work was not expected to be reliable or physically accurate.  Our $\log\,g_\star$ values determined using the MIST isochrones better represent the physical conditions of the star. A comparison of the these and the spectroscopic values is presented in Figure \ref{fig:logg_comp}.

This comparison is analogous to that by \citet{tsantaki2013}, who saw an underestimated when comparison their spectroscopic $\log\,g_\star$ values to those estimates from parallaxes. From Figure \ref{fig:logg_comp}, it is also apparent that the uncertainties on the MIST $\log\,g_\star$s are greatly reduced compared to the spectroscopic values. Taking into account these higher precisions, and the general consensus of the literature, we would suggest that any further analysis required surface gravities is done using the MIST $\log\,g_\star$ values.

The full set of stellar parameters is presented in an online machine-readable table.

\section{Conclusion}
\label{conclusions}

In this work we present the homogeneous stellar analysis of the BEBOP sample. With effective temperatures ranging from below 5000 K to over 7000 K, the BEBOP sample stretches the limit of what can be achieved with homogeneous analysis. Our spectroscopic method uses the EW method and spectral synthesis in succession to derive all stellar atmospheric parameters, but make use of the computational speed provided by the EW method. We provide a public tool, {\tt PAWS}, that we used to perform our analysis, but is applicable to any spectra of FGK dwarfs.

Our method has been demonstrated to produce reliable parameters for solar-type main sequence stars as was the aim during its production, with the advantage of using the EW method to generate intial parameters for spectral synthesis made clear. Although the input parameters do not have a significant impact on the final derived ones, beginning in a parameter space that is close to the physical one saves on computation time and allows more iterations to be dedicated to the refinement of well-fitting parameters.  Comparison to literature parameters for the BEBOP data set revealed strong agreement throughout this work.

Emphasis is put on a minimum SNR of 100, however in this analysis we find no targets deviate significantly from expected results purely due to low SNR despite 93 of 194 spectra not meeting this. The ability to reliably return stellar parameters for spectra of SNR $<$ 100 allows a wide range of spectra to be analysed that may not be possible to use the equivalent widths method for. As previously discussed in Section \ref{lowsnr}, EW analyses require high SNR spectra. Within the BEBOP sample, only 11\% of combined spectra reach an SNR of 200, indicating that a pure equivalent widths analysis of the sample would not be viable. Poor performance at low SNR indicates that the EW part of our analysis should be skipped for particularly low SNR targets, instead using only synthesis to provide stellar parameters. The use of the EW method in such a scenario is unlikely to initiate the synthesis method in a parameter space reflective of the physical situation. As a result, the time-saving aspect of the EW method would be obsolete here.

Stellar parameters for the BEBOP data set from literature are far from homogeneous. The majority of targets have stellar parameters publicly available from only Gaia DR3 and the TIC. Available parameters from these sources are limited, therefore a homogeneous picture of BEBOP targets cannot be created from current literature. In providing essential stellar parameters for every BEBOP target using the same method, with the same inputs, homogeneity is ensured for any further analysis of the BEBOP sample.

\section*{Data availability}
\label{data_avail}
All underlying data is available either in the appendix/online supporting material or will be available via VizieR at CDS.
All spectra analysed in this paper are either already publicly available on the OHP and ESO archives respectively or will be upon publication of this paper.
PAWS is available on GitHub\footnote{\url{https://github.com/alixviolet/PAWS}} along with our updated linemasks, and detailed in Appendix \ref{appendix:A}. 

\section*{Acknowledgments}
We are grateful to Sergi Blanco-Cuaresma for providing invaluable insights in his code.

We thank the anonymous referee for their insightful and valuable feedback that improved this work.

This paper's results were only possible with the collection of nearly 3000 spectra obtained at two observatories. We therefore particularly thank the staff at ESO's observatory of La Silla, and at the Observatoire de Haute-Provence (OHP) for their very kind attention, and the extra work they produced during the COVID pandemic, while travel restrictions prevented us from going to the telescopes. This paper is based on data obtained at ESO under proposals 106.212H.001, 106.212H.007, 1101.C-0721(A), 106.21TJ.001, 095.C-0367(A), 092.D-0261(A), 099.C-0138(B), 0101.C-0510(C), 099.C-0138(A), 105.20GX.001, 106.21ER.001, 108.22A8.001, 0101.C-0407(A) and is based on observations collected at OHP under proposals 19A.OPT.TRIA, 18B.PNP.SANT, 19A.PNP.SANT, 19B.PNP.SANT, 20A.PNP.SANT, 20B.PNP.SANT, 21A.PNP.SANT, 22A.PNP.SANT, 22B.PNP.SANT.

The authors would like to thank Maria Bergemann for her assistance with some of the spectroscopic analysis that allowed us to refine our methods, and her comment on an earlier version of this paper.

This research has made use of the Exoplanet Follow-up Observation Program (ExoFOP; DOI: 10.26134/ExoFOP5) website, which is operated by the California Institute of Technology, under contract with the National Aeronautics and Space Administration under the Exoplanet Exploration Program.

This work has made use of data from the European Space Agency (ESA) mission {\it Gaia} (\url{https://www.cosmos.esa.int/gaia}), processed by the {\it Gaia} Data Processing and Analysis Consortium (DPAC, \url{https://www.cosmos.esa.int/web/gaia/dpac/consortium}). Funding for the DPAC has been provided by national institutions, in particular the institutions
participating in the {\it Gaia} Multilateral Agreement.

This publication makes use of data products from the Wide-field Infrared Survey Explorer, which is a joint project of the University of California, Los Angeles, and the Jet Propulsion Laboratory/California Institute of Technology, funded by the National Aeronautics and Space Administration.

This research is supported from the European Research Council (ERC) under the European Union's Horizon 2020 research and innovation programme (grant agreement n$^\circ$ 803193/BEBOP), and by a Leverhulme Trust Research Project Grant (n$^\circ$ RPG-2018-418).

AVF acknowledges the support of the IOP through the Bell Burnell Graduate Scholarship Fund.

ACC acknowledges support from STFC consolidated grant numbers ST/R000824/1 and ST/V000861/1.

MRS acknowledges support from the UK Science and Technology Facilities Council (ST/T000295/1).

PM and NM were supported by STFC grant number ST/S001301/1.

E.M. acknowledges funding from FAPEMIG under project number APQ-02493-22
and research productivity grant number 309829/2022-4 awarded by the CNPq,
Brazil.

This project has received funding from the European Research Council (ERC) under the European Union's Horizon 2020 research and innovation programme (project {\sc Spice Dune}, grant agreement No 947634). 

DJA is supported by UKRI through the STFC (ST/R00384X/1) and EPSRC (EP/X027562/1).

J. R. French acknowledges the support of a University of Leicester College of Science and Engineering Studentship. The authors would like to thank Adam Green in his work towards the eclipsing binary sample.

Support for this work was provided by NASA through the NASA Hubble Fellowship grant \#HF2-51464 awarded by the Space Telescope Science Institute, which is operated by the Association of Universities for Research in Astronomy, Inc., for NASA, under contract NAS5-26555

A.C. acknowledge funding from the French ANR under contract number ANR\-18\-CE31\-0019 (SPlaSH).
This work is supported by the French National Research Agency in the framework of the Investissements d'Avenir program (ANR-15-IDEX-02), through the funding of the ``Origin of Life" project of the Grenoble-Alpes University.

EW, acknowledges support from the ERC Consolidator Grant funding scheme (project ASTEROCHRONOMETRY, G.A. n. 772293 \url{http://www.asterochronometry.eu})

IB thanks the support of the Programme National de Plan\'etologie (PNP) of CNRS-INSU co-funded by CNES.



\bibliographystyle{mnras}
\bibliography{bibliography} 




\appendix
\section{{\tt PAWS}}
\label{appendix:A}

To perform our homogeneous analysis, using several functionalities of iSpec, we wrote a python tool, {\tt PAWS}, to more easily analyse all the spectra. {\tt PAWS} stands for Parameters Approximated With Synthesis.

{\tt PAWS} is presented as an interactive GUI that can handle both individual spectra, and also perform bulk analysis. It is publicly available as described in Section \ref{paws}. Minimal inputs are required, namely a folder path, spectral resolution, desired wavelength range, and an initial estimate of spectral type. Input spectra can either be individual spectra from an instrument's reduction pipeline, or pre-combined. The GUI is versatile and suitable for a variety of scenarios, with a setup panel providing simple options to allow flexibility of input.

Spectra from differing instruments and observing modes must be sorted into separate folders by the user before input to {\tt PAWS}. Once {\tt PAWS} is initiated, individual spectra are corrected for RV to put them in the lab-frame and undergo continuum normalisation. Spectra meeting the SNR threshold of 20 established in Section \ref{lowsnr} after normalisation are co-added, with the resulting spectrum stored in the appropriate folder as `prepared\_spectrum.fits'. Initial values of T$\rm_{eff}$, $\log\,g_\star$, and metallicity are then generated by applying the EW method to this spectrum using the \textit{Run Analysis Stage 1} button in the {\tt PAWS} GUI. 
Upon activating \textit{Run Analysis} in the {\tt PAWS} GUI, the EW parameters are fed as inputs to the synthesis method, with metallicity fixed due to the proficiency of the EW method in using the iron lines. As $v_{\rm mic}$ cannot be determined via spectral synthesis, it was also necessary to fix this to the value determined as detailed in Section \ref{EW}. Results of the full analysis are then saved along with the original data. The final results folder includes the files detailed in Table \ref{files}.

{\tt PAWS} has been set up to offer more functionality than used for the analysis of the paper to allow users to make different choices. These changeable options include: 

\begin{itemize}

\item {\tt PAWS} can also use non-solar input parameters for the equivalent widths section, depending on initial knowledge of a target's spectral type.  As the majority of BEBOP targets were fitted with G type masks by instrumental data reduction, the solar inputs to the equivalent widths method were not changed during their analysis, further adding to the homogeneity of the method. However, allowing for future investigation of a wider variety of targets, K and F type input parameters have been added as an option for {\tt PAWS}.

\item Although recommended to keep this in place to avoid the introduction of contamination to the combined spectrum, the SNR filter can be removed when using {\tt PAWS} if necessary.

\item Depending on the format of the input spectra, it can be selected whether they need to be coadded, or to use a pre-coadded spectrum. Continuum normalisation will occur regardless.

\item The instrumental resolution and wavelength range can be customised, to allow for not only the correct resolution to be used, but also to ensure only the desired section of the spectrum is analysed.
\end{itemize}

\begin{table*}
    
    \caption{Files saved by {\tt PAWS} into the target folders and their descriptions.}
    
    \begin{tabular}{ll}
    \hline
        File Name & Description \\
        \hline
        \textit{SNR\_list.txt} & Text file listing the SNR of all spectra used for coadding, with the last value being the co-added spectrum SNR. \\
        \textit{prepared\_1\_.fits} & FITS file of the co-added spectrum as flux vs wavelength (nm), including the flux errors. \\
        \textit{params\_EW.csv} & Table of parameters and errors derived only by the equivalent widths method (i.e. those used as synthesis inputs).\\
        \textit{used\_linemasks.csv} & Table of the wavelengths and properties of the iron lines used by the equivalent widths method.\\
        \textit{params\_synth\_pipeline.csv} & Table of the final parameters derived by the pipeline.\\
        \textit{fitted\_line\_params.csv} & Table of the wavelengths and properties of the lines used in synthesis fitted.\\
        \textit{synthesized\_spectrum\_pipeline.fits} & FITS file of the synthesized spectrum as flux vs wavelength (nm). \\
        \hline
    \end{tabular}
    
    \label{files}
\end{table*}

\section{Additional tables}

Table \ref{tab:lines} presents the lines that were removed from the standard iSpec linelist in our analysis.

\begin{table}
    \centering
    \caption{Lines removed from analysis due to poor chi-squared results from SOPHIE HE mode synthesis fitting.}
    \begin{tabular}{|l|l|l|}
    \hline
        wave\_peak (nm) & wave\_base (nm) & wave\_top (nm) \\ \hline
        480.0649 & 480.061999995742 & 480.064999995742 \\ 
        500.5712 & 500.570999995257 & 500.572999995257 \\ 
        502.9618 & 502.960999995201 & 502.9629999952 \\ 
        503.6922 & 503.690999995183 & 503.693999995183 \\
        510.403 & 510.402999995025 & 510.405999995025 \\ 
        531.0686 & 531.055 & 531.089 \\ 
        531.2856 & 531.271 & 531.306 \\ 
        531.7525 & 531.73 & 531.782 \\ 
        531.8771 & 531.856 & 531.892 \\ 
        531.9035 & 531.892 & 531.913 \\ 
        532.0036 & 531.991 & 532.026 \\ 
        532.4179 & 532.305 & 532.512 \\ 
        532.5552 & 532.538 & 532.58 \\ 
        532.7252 & 532.705 & 532.737 \\ 
        532.9138 & 532.89 & 532.951 \\ 
        532.9784 & 532.951 & 532.986 \\ 
        532.9989 & 532.986 & 533.026 \\ 
        534.0447 & 534.042999994466 & 534.045999994466 \\ 
        539.7618 & 539.75899999433 & 539.76199999433 \\ 
        547.2709 & 547.267999994153 & 547.270999994153 \\ 
        549.1832 & 549.182999994108 & 549.185999994108 \\ 
        563.8262 & 563.823999993761 & 563.826999993761 \\ 
        566.9943 & 566.993999993686 & 566.995999993686 \\ 
        578.7919 & 578.789999993407 & 578.792999993407 \\ 
        608.5258 & 608.523999992704 & 608.526999992704 \\ 
        623.0722 & 623.07199999236 & 623.07499999236 \\ 
        675.2707 & 675.268999991126 & 675.270999991126 \\ 
        \hline
    \end{tabular}
    \label{tab:lines}
\end{table}

\section{Synthesis-Only Targets}
\label{speedymethod}

\begin{table}
    \centering
    \caption{Targets for which the EW step was skipped during analysis, categorised by the reasoning for doing so.}
    \begin{tabular}{p{0.4\textwidth}p{0.5\textwidth}}
        \hline
         Reason for skipping EW & Targets\\
         \hline
         Low SNR (<50) & J1916-04, J0709-48, J1246-48, J1126-55, J1757+32, J0211+36, J2303+39, J1623+05, J0641+28, J1558+24\\
         \hline
         $v\sin i_\star\gtrsim 5 ~\rm km\,s^{-1}$& J1008-29, J0456-74, J1540-09, J0412-44, J0525-55, J0407-23, J0759-69, J0610-56, J1341-30, J0432-33, J1258-58, J0700+09, J0719+25, J0601+37, J0114+54, J1836+27, J0702+42, J0719+10\\

    \end{tabular}
    \label{tab:skippy}
\end{table}

\bsp	
\label{lastpage}
\end{document}